\documentclass[fleqn,usenatbib]{mnras}

\usepackage{newtxtext,newtxmath}

\usepackage[T1]{fontenc}

\DeclareRobustCommand{\VAN}[3]{#2}
\let\VANthebibliography\thebibliography
\def\thebibliography{\DeclareRobustCommand{\VAN}[3]{##3}\VANthebibliography}

\usepackage{xcolor}

\usepackage{graphicx}	
\usepackage{amsmath}	

\title[WFST GC]{Tidal structures of six globular clusters from the Wide Field Survey Telescope (WFST) pilot survey}

\author[Zhen W., et al.]{Zhen Wan$^{1,2}$\thanks{E-mail: zhen\_wan@ustc.edu.cn}, Lulu Fan$^{1,2,5}$, Xuzhi Li$^{3,4}$, Xu Kong$^{1,2,5}$, Tinggui Wang$^{1,2,5}$, Qingfeng Zhu$^{1,2,5}$, Ji-an Jiang$^{1,2,6}$,\newauthor
Minxuan Cai$^{1,2}$,Zelin Xu$^{1,2}$, Xianzhong Zheng$^{7}$, Jingquan Cheng$^{8}$, Feng Li$^{9}$, Ming Liang$^{10}$, Hao Liu$^{9}$, \newauthor
Wentao Luo$^{5}$, Jinlong Tang$^{11}$, Hairen Wang$^{8}$, Jian Wang$^{9,5}$, Yongquan Xue$^{1,2}$, Dazhi Yao$^{8}$, Hongfei Zhang$^{9}$, \newauthor
Wen Zhao$^{1,2}$
\\
$^{1}$Department of Astronomy, University of Science and Technology of China, Hefei 230026, People's Republic of China\\
$^{2}$School of Astronomy and Space Science, University of Science and Technology of China, Hefei 230026, China\\
$^{3}$School of Mathematics and Physics, Anqing Normal University, Anqing 246133, China\\
$^{4}$Institute of Astronomy and Astrophysics, Anqing Normal University, Anqing, Anhui, 246133, People’s Republic of China\\
$^{5}$Institute of Deep Space Sciences, Deep Space Exploration Laboratory, Hefei 230026, China\\
$^{6}$National Astronomical Observatory of Japan, National Institutes of Natural Sciences, Tokyo 181-8588, Japan\\
$^{7}$Tsung-Dao Lee Institute and Key Laboratory for Particle Physics, Astrophysics and Cosmology, Ministry of Education, Shanghai Jiao Tong University, \\$^{\ \ }$Shanghai, 201210, China\\
$^{8}$Purple Mountain Observatory, Chinese Academy of Sciences, Nanjing 210023, China\\
$^{9}$State Key Laboratory of Particle Detection and Electronics, University of Science and Technology of China, Hefei 230026, China\\
$^{10}$National Optical Astronomy Observatory (NSF’s National Optical-Infrared Astronomy Research Laboratory) 950 N Cherry Ave. Tucson Arizona 85726, USA\\
$^{11}$Institute of Optics and Electronics, Chinese Academy of Sciences, Chengdu 610209, China\\
}

\date{Accepted XXX. Received YYY; in original form ZZZ}

\pubyear{\the\year{}}

\begin{document}
\label{firstpage}
\pagerange{\pageref{firstpage}--\pageref{lastpage}}
\maketitle

\begin{abstract}
We carry out an imaging survey of six globular clusters (GCs) with a limit magnitude to $22\ \mathrm{mag}$ at the $5\sigma$ level, down to the main sequence stars of the respective cluster, as one of the pilot observing program of the Wide Field Survey Telescope (WFST). 
This paper present the early results of this survey, where we investigate the tidal characters at the periphery of the clusters NGC~4147, NGC~5024, NGC~5053, NGC~5272, NGC~5904 and NGC~6341. We present the estimated number density of cluster candidates and their spatial distribution. We confirm the presence of tidal arms in NGC~4147 and NGC~5904 and identify several intriguing potential tidal structures in NGC~4147, NGC~5024, NGC~5272, corroborated the elliptical morphology of the periphery of NGC 6341. WFST shows its ability to detect faint main-sequence stars of clusters beyond 15 kpc in helio-centric distance. Our findings underscore WFST's capability for probing faint structural features in GCs, paving the  way for future in-depth studies, especially for the search of the large scale tidal streams associated with the clusters with the future wide field survey.  

\end{abstract}

\begin{keywords}
surveys -- globular clusters: general -- Galaxy: structure
\end{keywords}



\section{Introduction}
The evolution of galaxies is fundamentally a history of accretion and merger events, which contribute to the growth of the Milky Way by supplying material \citep[e.g.,][]{1994Natur.370..194I}. Alongside these significant merger events, the GC systems of the progenitor galaxies are also assimilated into the Milky Way \citep[e.g.,][]{2019MNRAS.486.3180K,2020ApJ...889..107C}. By their intrinsic nature, GCs preserve the environmental conditions of their birthplaces and record the interaction history during mergers, offering a unique window into the properties of progenitor galaxies and the merger history. Strong tidal interactions with the Milky Way's gravitational potential reshape these GCs, leading to deformed stellar envelopes and the formation of extended tidal tails in some cases \citep[see ][ and references therein]{2021ApJ...914..123I,2022MNRAS.tmp..731Z}. The spatial distribution of these tidal tails provides critical constraints on the properties of the Milky Way's gravitational potential, particularly when compared with $N$-body simulations \citep[e.g.,][]{2010ApJ...714..229L}.

The connection between GCs, stellar streams, and their surrounding tidal tails or features \citep[e.g.,][]{2021ApJ...914..123I} reveals that GCs are not simple, spherically symmetric structures densely packed with stars. Instead, their member stars can escape to significant distances, and the clusters themselves are dynamically reshaped by the Milky Way's gravitational potential. Notably, tidal features associated with GCs can exhibit large scale asymmetry \citep[e.g.,][]{2012MNRAS.419...14C,2014MNRAS.445.2971C}, underscoring the need for an unbiased study of each single cluster. Such a study requires a survey capable of providing deep photometry and a wide field of view \citep[e.g.,][]{2024A&A...683A.151P, 2025MNRAS.537.1586P}, which are essential for a systematic understanding of GCs system. Furthermore, the morphology and spatial distribution of tidal features, combined with robust candidate selection based on photometric data, are critical for enabling future chemo-dynamical studies. These studies will rely on spectroscopic follow-up observations, which are often resource-intensive and require careful target prioritization \citep[e.g.,][]{2019MNRAS.490.3508L,2025A&A...693A..69A}.

Pioneering photometric surveys have provided significant advantages for studying the stellar populations and spatial distributions of GCs \citep[e.g.,][]{2003A&A...405..577B, 2011MNRAS.416..393B}. However, the faint outer regions of these clusters have only recently become more accessible and scientifically significant, thanks to advancements in observational techniques and instrumentation \citep[e.g.,][]{2019ApJ...887L..12B, 2022MNRAS.tmp..731Z, 2024AJ....167..279P, 2024A&A...683A.151P}. Mapping the faint structures surrounding the Milky Way remains a key focus for both existing and upcoming optical surveys, such as the Dark Energy Survey (DES), the Large Synoptic Survey Telescope (LSST), the Euclid mission, the Chinese Space Station Telescope (CSST), and the James Webb Space Telescope (JWST) \citep[e.g.,][]{2005astro.ph.10346T, 2009arXiv0912.0201L, 2024arXiv240513491E, 2018cosp...42E3821Z, 2006SSRv..123..485G}.

The Wide Field Survey Telescope (WFST)\footnote{https://wfst.ustc.edu.cn} is one of the latest optical survey telescopes to commence operations \citep{2023SCPMA..6609512W}. It features a 2.5-meter primary mirror and is equipped with five corrector lenses, an atmospheric dispersion compensator (ADC), and active optics (AO). The telescope's filter system includes five bands (ugriz) and a white band. Its scientific imaging array comprises nine CCDs, each with $\mathrm{9k\times9k}$ pixels, enabling a field of view of 3 degrees. WFST is designed to survey the northern sky with unprecedented sensitivity, aiming to explore the dynamic universe and capture time-domain phenomena. Key scientific objectives include the detection of supernovae \citep[SNe,][]{2023MNRAS.525..246H}, tidal disruption events \citep[TDEs,][]{2022MNRAS.513.2422L}, variables \citep{2024arXiv241212601L},  optical counterparts of gravitational wave events \citep{2023ApJ...947...59L}, active galactic nuclei \citep[AGN,][]{2024ApJ...976..155S} variability, and solar system objects \citep[SSOs,][]{2025arXiv250112460L, 2025arXiv250117472W}.

Since the establishment of the Lenghu Observatory \citep{2021Natur.596..353D}, WFST represents the first large optical survey telescope to commence its mission at this new site. As one of its primary scientific objectives, WFST will conduct a wide-field survey covering approximately $8000\ \mathrm{deg^2}$ in the northern sky, reaching a depth of over 24 magnitudes. This deep dataset will facilitate systematic studies of large faint structures around the Galaxy, including the Galactic halo structure, faint dwarf galaxies, and the outer regions and the tidal features of Galactic GCs, which will contribute to near-field cosmology research and enhance our understanding of small-scale structures. Against this backdrop, we have initiated a study of the northern sky GC system using WFST. This project aims to survey northern sky GCs to depths reaching their main sequences, with the goals of investigating their morphology and traces of tidal interactions, identifying short-period variable stars within the clusters, and analysing stellar populations through the advantage of the future deep $u$-band photometry. In this work, we present the first results from WFST observations of GCs from the pilot survey, focusing on their structural properties, particularly the tidal structures in their outer regions.

This paper is organized as follows: Sec.~\ref{sec:survey_summary} provides an overview of the observations; Sec.~\ref{sec:data_reduction} illustrate the data reduction process; the observational results are presented and discussed in Sec.~\ref{sec:results}; and the conclusions of this study are summarized in Sec.~\ref{sec:conclusion}.

\begin{figure*}
    \centering
    \includegraphics[width=0.9\linewidth]{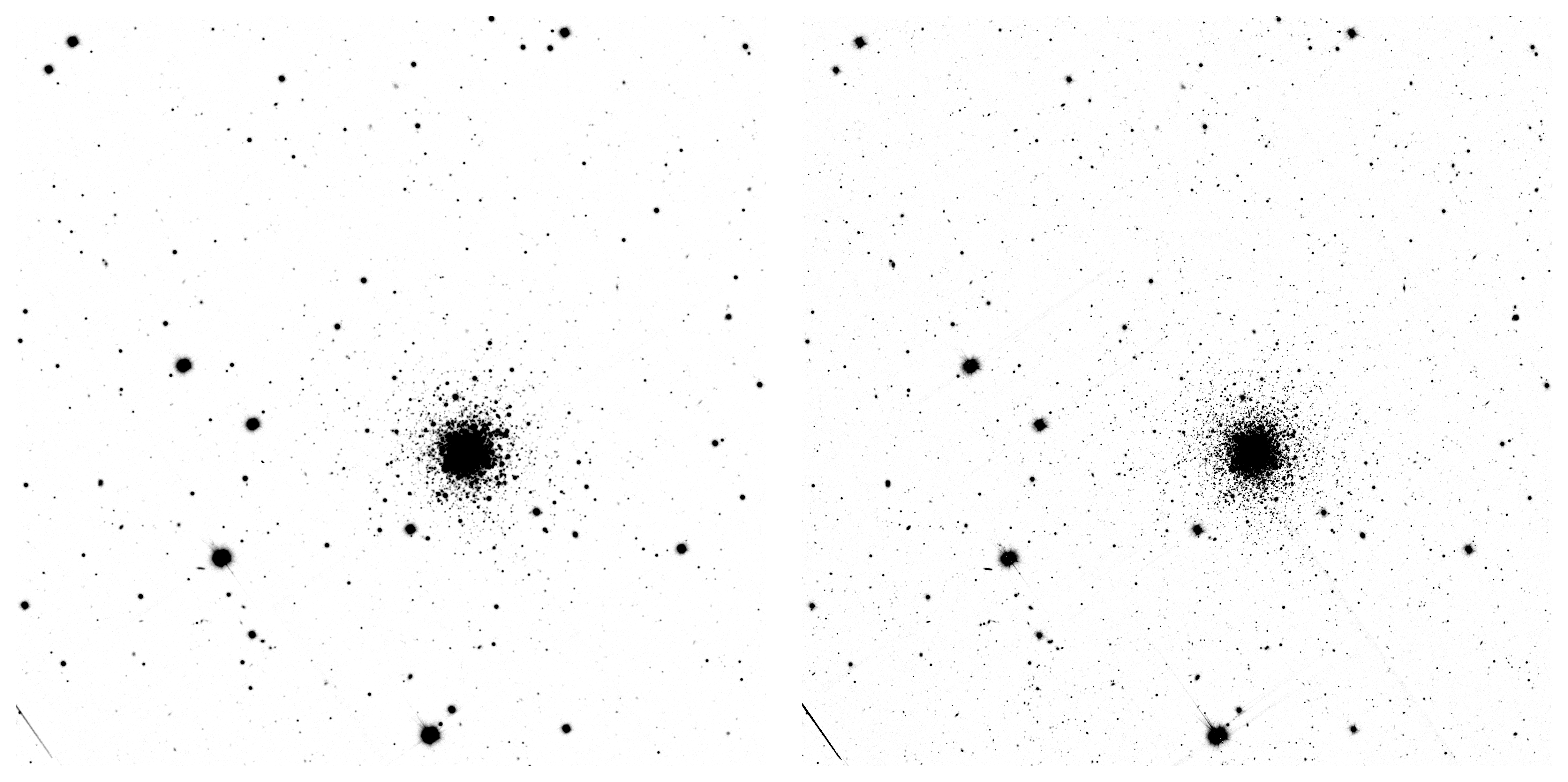}
    \caption{The stacked {\it g} (left) and {\it r} (right) images of central $\sim 100^{\prime\prime}\times100^{\prime\prime}$ of NGC~4147 are presented. The {\it r}-band exposure exhibits greater depth, primarily attributable to more favorable weather conditions during observation.}
    \label{fig:NGC4147_exposure}
\end{figure*}

\section{Survey summary}
\label{sec:survey_summary}

During the pilot survey conducted from March to July 2024, we performed a series of single-pointing observations of GCs. Tab.~\ref{tab:cluster_summary} summarise the fundamental properties of the GCs in our observations. Exposures pointing towards 5 globular clusters (NGC~5024 and NGC~5053 are contained in the single pointing) in two bands ({\it g} and {\it r}) were obtained, with the total exposure time for each target designed to ensure the detection of members below the main-sequence turnoff. In total, 214 exposures with the total exposure time of 8040 seconds were collected, and the observational details are summarized in Tab~\ref{tab:observation_summary}.

The telescope's large field of view (FOV) enables a single exposure to cover a 3-degree area, which, for a cluster located 10 kpc away, corresponds to a physical radius of approximately 500 pc around the cluster. This extensive FOV not only ensures efficient survey coverage within the limited time available during the pilot phase but also facilitates the study of extended features associated with the clusters.

\begin{table*}
    \centering
    \begin{tabular}{c|c|c|c|c|c|c}
       Cluster & Coordinate & $R_{\odot}$ ($\mathrm{kpc}$) & $R_{GC}$ ($\mathrm{kpc}$) & Proper motion ($\mathrm{mas/yr}$) & $V_{los}$ ($\mathrm{km/s}$)\\
       \hline 
        NGC~4147 & $(182^{\circ}.5262,18^{\circ}.5426)$ & $18.54 \pm 0.21$ & $20.74 \pm 0.19$ & $(-1.690 \pm 0.014, -2.092 \pm 0.013)$ & $179.35 \pm 0.31$\\
        NGC~5024 & $(198^{\circ}.2302, 18^{\circ}.1681)$ & $18.50 \pm 0.18$ & $19.00 \pm 0.16$& $(-0.131 \pm 0.005,-1.332 \pm 0.005)$ & $-63.37 \pm 0.25$\\
        NGC~5053 & $(199^{\circ}.1128,17^{\circ}.7002)$ & $17.54 \pm 0.23$ & $18.01 \pm 0.20$ & $(-0.338 \pm 0.007, -1.214 \pm 0.007)$ & $42.82 \pm 0.25$ \\ 
        NGC~5272 &$(205^{\circ}.5484, 28^{\circ}.3772)$& $10.18\pm0.08$& $12.09\pm.06$&$(-0.153\pm0.007,-2.679\pm0.007)$ & $-147.20\pm0.27$\\
        NGC~5904 &$(229^{\circ}.6384,2^{\circ}.0810)$&$7.48\pm0.06$&$6.27\pm0.02$&$(4.073\pm0.006, -9.872\pm0.006)$&$53.50\pm0.25$\\
        NGC~6341 & $(259^{\circ}.2807,43^{\circ}.1359)$ & $8.50\pm0.07$&$9.85\pm0.04$&$(-4.934\pm0.012, -0.630\pm0.012)$&$-120.55\pm0.27$\\
        \hline 
    \end{tabular}
    \caption{The basic properties of the GCs we concern, which are adopted from \citet{2021MNRAS.505.5957B}. The first column is the identifier of the cluster; the second column is the coordinate of the cluster in the Equatorial Coordinate System; the third column is the heliocentric distance; the fourth column is the Galactocentric distance; the fifth column is the proper motion and the last column is the line-of-sight (LOS) velocity. }
    \label{tab:cluster_summary}
\end{table*}

\begin{table}
    \centering
    \begin{tabular}{c|c|c|c|c}
       Cluster & Filter & Exp. time (s) & Epoch & FWHM ($^{\prime\prime}$)\\
       \hline 
        NGC~4147 & g/r &  60/60 & 12/15 & $\sim 2.52/1.36$\\
        NGC~5024 & g/r & 60/60 & 12/15 & $\sim 2.05/1.24$\\
        NGC~5272 & g/r & 30/30 & 25/25 & $\sim 1.59/1.18$\\
        NGC~5904 & g/r &  30/30 & 25/25 & $\sim 1.74/1.35$ \\
        NGC~6341 & g/r & 30/30 & 30/30 & $\sim 1.37/1.70$ \\
        \hline 
    \end{tabular}
    \caption{Summary of the GC pilot survey. The first column shows the name of the cluster; the second column is the filter name of each set of exposures; the third column shows the single exposures times; the fourth column shows the observation epochs numbers and the fifth column shows the FWHM of the final stacked images}
    \label{tab:observation_summary}
\end{table}

The general design of the exposure time is to reach a depth that covers a significant portion of the main sequence turn-off. The actual photometric limit, however, is depended on the weather as well. For example, Fig.~\ref{fig:NGC4147_exposure} shows the stacked images of NGC~4147 in two bands. The clearly deeper in r band is primarily due to a much better seeing condition.

\section{Data reduction}
\label{sec:data_reduction}

The raw data collected by WFST require further calibration, which includes standard calibration exposures taken routinely each night. These consist of 20–30 bias exposures acquired before and after observations, as well as 10–20 flat exposures obtained during twilight, depending on the required filters and weather conditions. The calibration exposures undergo preprocessing as follows. First, the bias frames are combined to create a "master bias," where each pixel value is the median of the corresponding pixels in the individual bias exposures. This master bias is then subtracted from all subsequent exposures. Similarly, the flat frames are combined to produce a "master flat." Each flat frame is normalized to its average count before the pixel values are median-combined to generate the master flat. The science exposures are then calibrated by subtracting the master bias and correcting for pixel-to-pixel response variations using the master flat. A more detailed description of the data reduction process can be found in \citet{2025arXiv250115018C}.

Further photometry is performed on the calibrated exposures primarily using the {\sc photutils} \citep{larry_bradley_2024_13989456} package. We begin with a preliminary source detection to generate an initial source catalog, which is also subsequently used for background estimation. A mask is created to mark regions with positive detections and a 10-pixel buffer around them. The background is estimated using a gridded approach with $200\times200$ pixel boxes and a kernel size of 2, employing the sigma-clipped {\sc Background2D} method to produce a low-resolution background map. For data from the $4^{th}$ CCD, we exclude a central $1^{\circ}\times1^{\circ}$ region where the cluster is located to prevent overestimation of the background level. The final background maps are derived by interpolating the low-resolution map, which is then subtracted from the science exposures to produce clean science images. Since the background of the globular cluster is estimated through the interpolation of the surrounding data, where the small scale fluctuation can leads to the uncertainties mainly influence the faint stars photometry. We estimate the background fluctuation in $3000\ \mathrm{pix} \times 3000\ \mathrm{pix}$ scale to be 2-3 ADU, which is significantly smaller than the background level of about 600 ADU, depending on the weather and moon distance.

To characterize the images, we perform deeper source detection and point spread function (PSF) measurements. Using {\sc DAOStarFinder}, we select sources with $roundness1^2 + roundness2^2 < 0.3$ as reliable detections. Here, $roundness1$ quantifies the ratio of the object’s bilateral (2-fold) to four-fold symmetry, while $roundness2$ measures the difference in the height of the best-fitting Gaussian function in the x-direction minus that in the y-direction, normalized by their average. A perfectly circular source would have both $roundness$ values equal to 0. The selected sources are crossmatched with the Pan-STARRS DR2 reference catalog \citep{2020ApJS..251....7F} within a magnitude range of 16 to 20 to ensure genuine detections. These matched stars are used to estimate the PSF of each exposure using the {\sc CircularGaussianPRF} model, which provides measurements of position, flux, and full-width half maximum (FWHM). The model is fitted to $15\times15$ pixel cutouts of the sources to derive their FWHM, and the PSF for each exposure is calculated as the sigma-clipped mean of the individual source measurements. This process is repeated for each CCD to account for slight variations in FWHM across the focal plane.

We perform PSF photometry using the {\sc PSFPhotometry} tool, with a fixed FWHM derived from the exposure. This enables the measurement of source positions, PSF fluxes, and the quality of photometry for detected point sources. To identify well-measured targets, we utilize the quality metrics $\mathrm{qfit}$, $\mathrm{cfit}$ and $\mathrm{flags}$ provided by {\sc PSFPhotometry}, applying the following selection criteria to select PSF measurement candidates:
\begin{gather}
    -0.1 < \mathrm{cfit} < 0.1 \notag \\
    \mathrm{qfit} < 0.1 \notag \\
    \mathrm{flags} = 0 \notag \\
    \mathrm{mag}_{err} < 0.1
    \label{eq:quality_selection}
\end{gather}

The parameter $\mathrm{cfit}$ is defined as the fit residual of the initial central pixel value divided by the fit flux, while $\mathrm{qfit}$ is defined as the absolute value of the sum of the fit residuals divided by the fit flux. Furthermore, the condition $\mathrm{flags} = 0$ ensures that the selected sources are not located within any mask or near the edge of the image. The flux selection criteria are applied to exclude sources that are either too faint or too bright, and the magnitude uncertainty constraints are used to select stars with a signal-to-noise ratio (SNR) of approximately 10. Additionally, we restrict the sample to stars with $2.5 < \mathrm{log}_{10}(\mathrm{flux}) < 5$ to avoid flux saturation or excessively faint measurements. The well-measured sources are crossmatched with the Pan-STARRS DR2 catalog (using the {\it g} or {\it r} bands, as appropriate) to calibrate the photometric zero point. They are also crossmatched with the Gaia DR3 catalog \citep{2023A&A...674A...1G} to calibrate the astrometric positions and the World Coordinate System (WCS) of the exposure.

The individual calibrated exposures are stacked to produce a deep co-added image. For each target, the exposures are combined using {\sc SWarp} \citep{2002ASPC..281..228B}, with the pixel variance propagated through the data calibration steps serving as the weight maps. Fig.~\ref{fig:NGC4147_exposure} presents the stacked images of NGC~4147 in the {\it g} (left) and {\it r} (right) bands. We emphasize that the final limiting magnitude is determined by combining the photometric data from both bands. The same photometric procedures are applied to the stacked images in the {\it g} and {\it r} bands separately. Subsequently, the multi-band data are crossmatched and merged to generate final source catalogs. In this study, we primarily utilize the deep photometric data derived from the stacked images, with a focus on analyzing the spatial distribution of the clusters.

\begin{figure}
    \centering
    \includegraphics[width=\columnwidth]{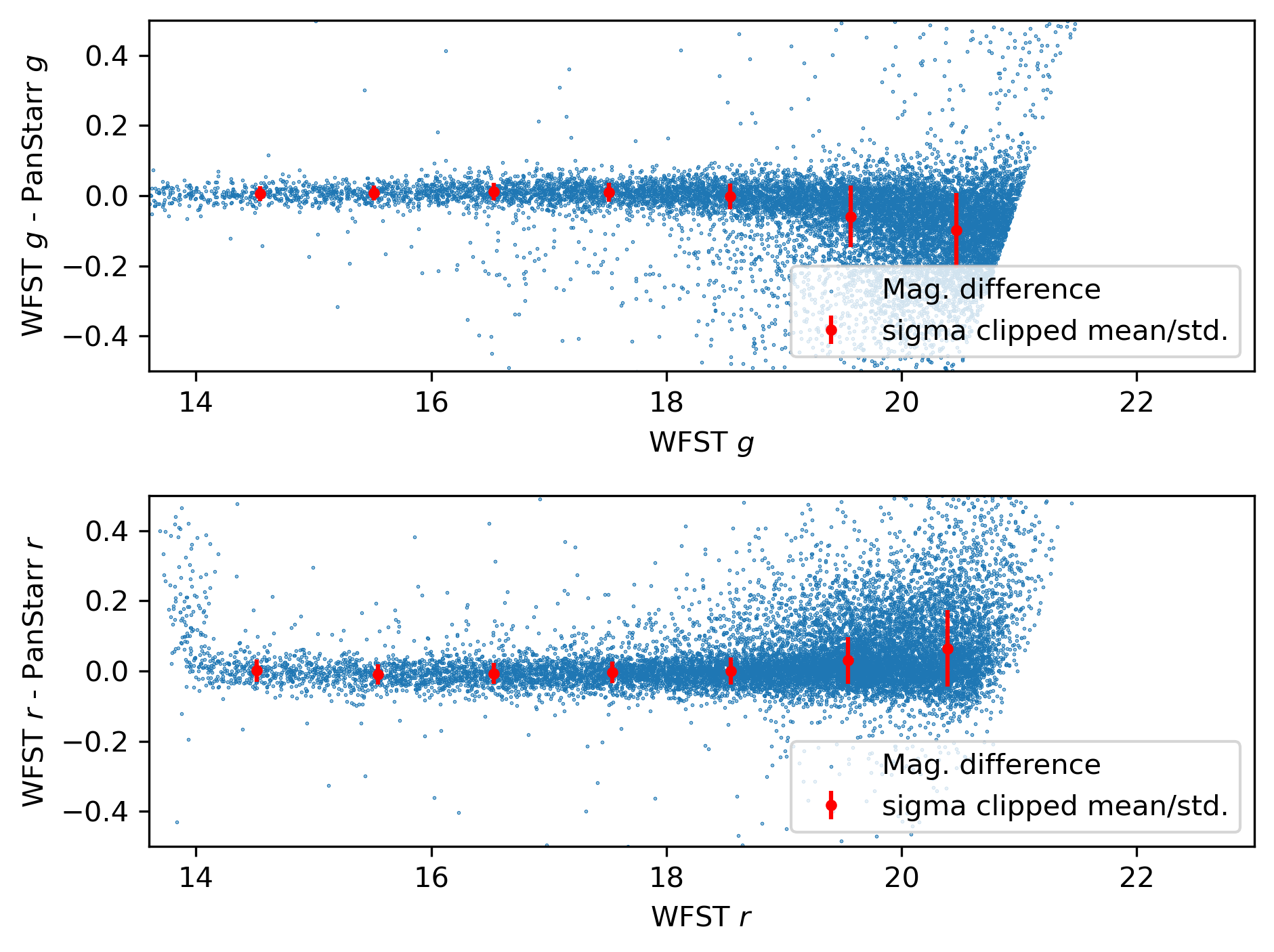}
    \caption{The magnitude difference between the WFST photometry and the reference Pan-Starrs catalog in $g$ (top) and $r$ (bottom) bands. The red errorbars shows the median and the standard deviation of the magnitude difference in each bin.}
    \label{fig:photometric_uncertainty}
\end{figure}

Fig.~\ref{fig:photometric_uncertainty} illustrates a comparison between our photometry results and those from the Pan-STARRS DR2 catalog. Our results demonstrate good consistency with the reference catalog. We note that the standard deviation of the magnitude differences presented in Fig.~\ref{fig:photometric_uncertainty} is larger than the intrinsic photometric uncertainties. This is because the standard deviation incorporates not only the statistical uncertainties from both catalogs but also the systematic uncertainties arising from differences between the two photometric systems.

\subsection{Cluster members selection}

We construct the colour-magnitude diagram (CMD) based on the two-band photometry restuls. Cluster members are selected from the source catalog based on the CMD. A widely adopted approach for identifying cluster members is the matched-filter method, which utilizes theoretical isochrones corresponding to the cluster's age and metallicity to determine its location in the CMD, and further the weight of each targets in the selection process. However, theoretical isochrones tailored to the WFST photometric system are currently unavailable. Given that the cluster's signature is most prominent in the central region, we employ an algorithm that integrates clustering and classification techniques to identify the cluster members' location in the CMD. This approach achieves an effect comparable to that of the matched-filter method.

First, we select stars within a radius of $0.1^{\circ}$ around the cluster center, where cluster members are expected to dominate. To further reduce contamination from field stars, we utilize proper motion and parallax data from Gaia DR3. Following \citet{2021A&A...649A...5F}, the Gaia dataset are cleaned with following quality controls: ${\it ruwe} < 1.4$, ${\it ipd\_frac\_multi\_peak} \le 2$ and ${\it ipd\_gof\_harmonic\_amplitude} < 0.1$. By combining the systemic proper motion and distance of the cluster from \citet{2021MNRAS.505.5978V}, we exclude stars whose parallax and proper motion measurements significantly deviate from the cluster's values. It is important to note that the astrometric information is used solely to eliminate stars that are clearly identified as field stars, thereby maximizing the signal from the cluster. However, most faint and distant stars—including both cluster members and field stars—lack reliable astrometric measurements and are therefore retained in the catalog. Specifically, we exclude stars follows any of the selections below:
\begin{gather}
    \omega > 0 \notag \\
    |\omega - \omega_{sys}| > \omega_{err}  \notag \\
    \sqrt{(\mu_{\alpha} - \mu_{\alpha0})^2 + (\mu_{\delta} - \mu_{\delta0})^2} > 4\ \mathrm{mas/yr}
\end{gather}
where $\omega$ and $\omega_{err}$ represent the parallax and parallax uncertainties from Gaia DR3, and $\omega_{sys}$ denotes the systemic parallax of the cluster derived from distance measurements. The terms $\mu_{\alpha}$ and $\mu_{\delta}$ correspond to the measured proper motion in the equatorial coordinate directions, while $\mu_{\alpha0}$ and $\mu_{\delta0}$ represent the systemic proper motion of the cluster. We note that the $4\ \mathrm{mas/yr}$ corresponds to $\approx 200\ \mathrm{km/s}$ for targets locates at $10\ \mathrm{kpc}$, which is significantly large enough for a confident non-member exclusion. Quality control from Eq.~\ref{eq:quality_selection} is also applied to the selected members. 
To further clean the cluster members, we apply the clustering algorithm HDBSCAN \citep{scikit-learn} to the samples from the central region. This method effectively selects stars located in the region of the CMD that are densely populated by cluster members. Specifically, it isolates stars predominantly on the main-sequence branch and further reduce the field contamination, establishing them as the fiducial sample of cluster members.

\begin{figure*}
    \centering
    \includegraphics[width=0.9\linewidth]{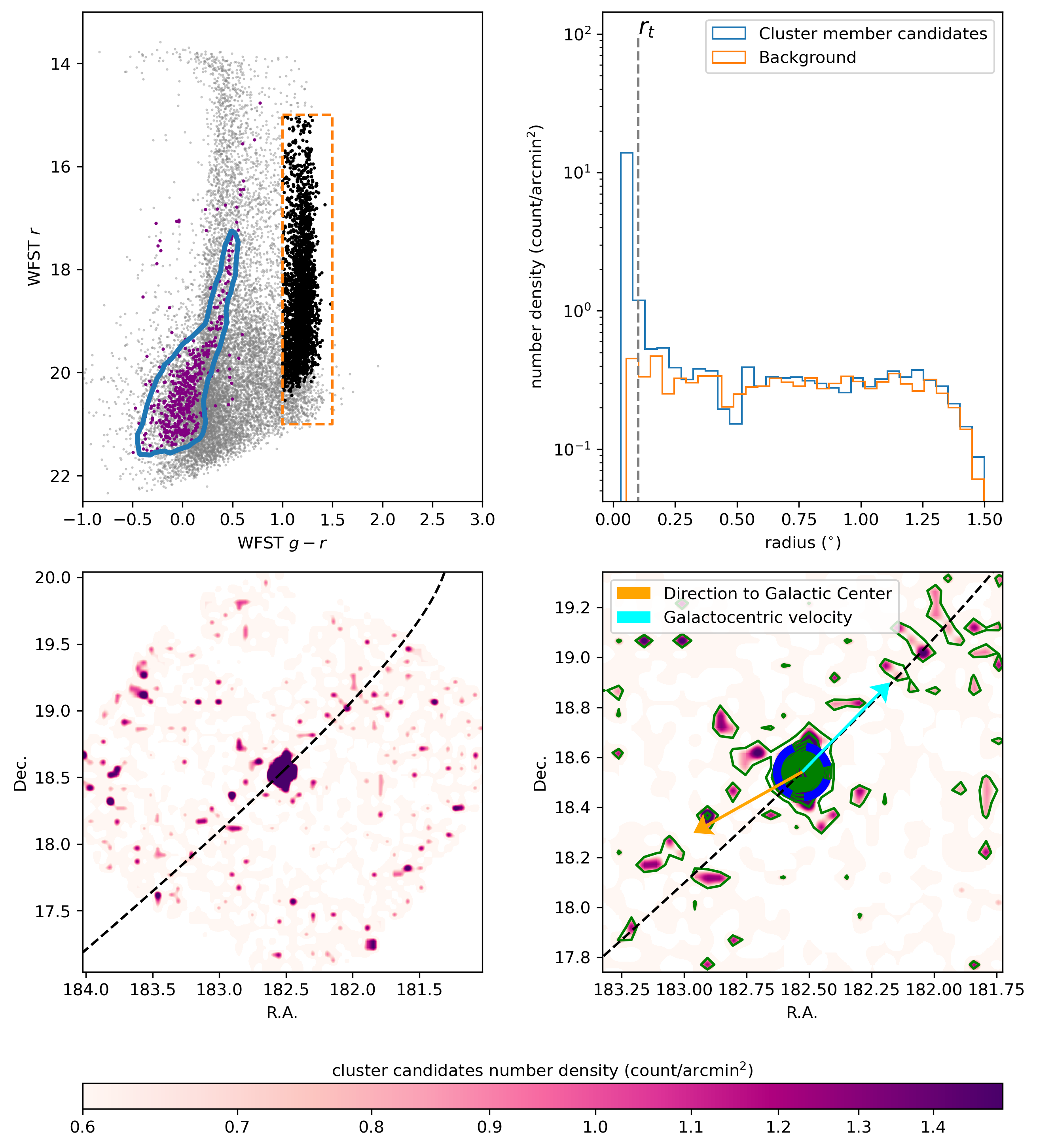}
    \caption{Top left: The color magnitude diagram of NGC~4147. The purple point are the stars within 0.1 degrees around the cluster center that are selected as a fiducial sample of the cluster member. The blue line outlines the regionwhere we adopt a stars to be a cluster member candidate calculated based on the distance to the fiducial  of the cluster members. The gray and the black points are the remaining background stars, while the black points within the red dashed rectangle are selected as the background level indicators.  Top right: the number density of the selected cluster candidate members (blue) and the background level (orange). The distribution are scaled to properly match the background level of the cluster candidates sample. The gray dashed line indicate the location of the tidal radius of the corresponding cluster. Bottom left: the number density distributions of the selected cluster candidates with the scaled background subtracted. Bottom right: the zoom-in view of the candidate distribution, where the blue dashed circle indicate the tidal radius of the cluster. We mark the direction of the Galactic centre as the orange arrow, and the Galacto-centric velocity of the cluster as the cyan arrow.}
    \label{fig:NGC4147_tails}
\end{figure*}

\begin{figure*}
    \centering
    \includegraphics[width=0.9\linewidth]{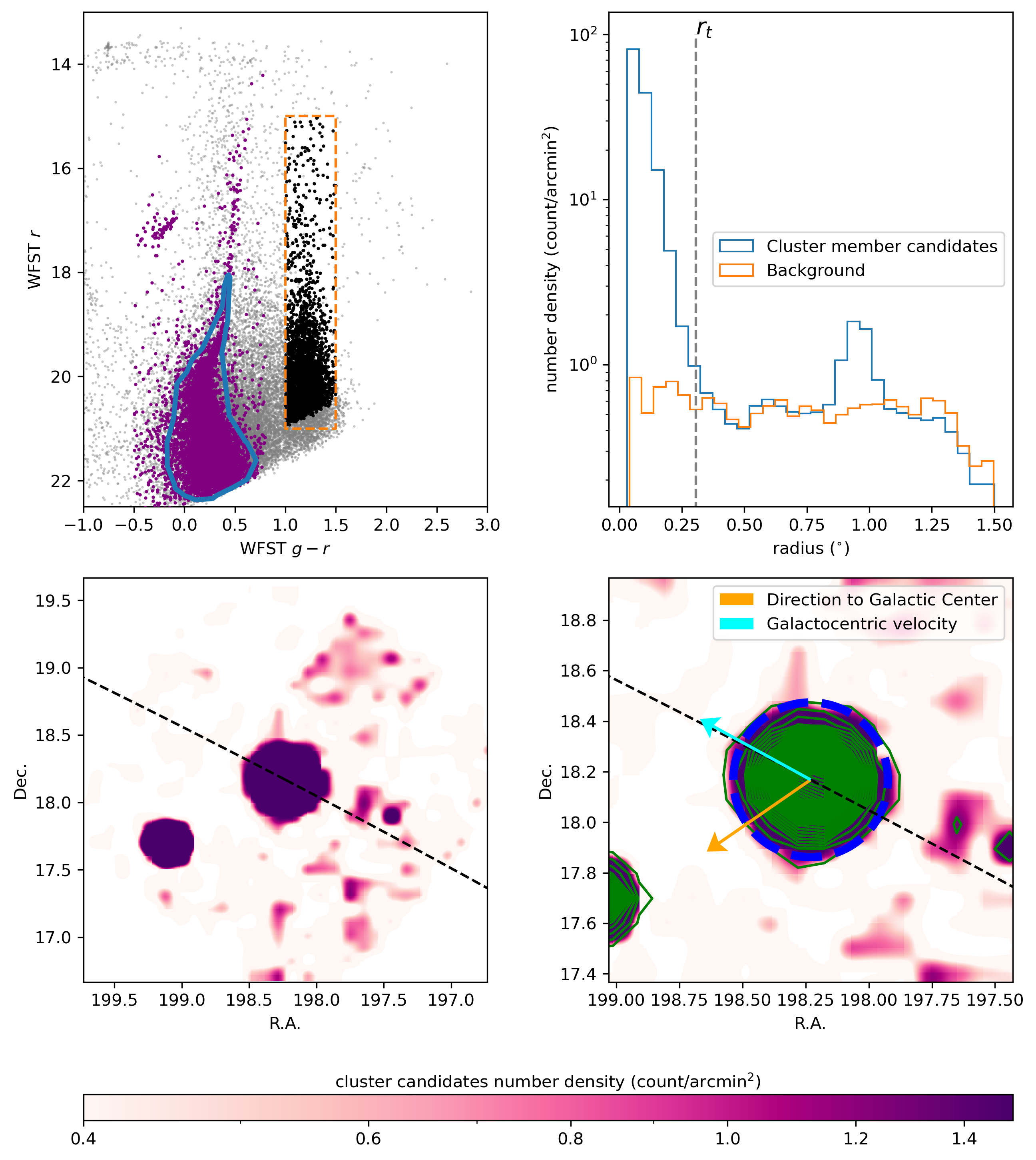}
    \caption{The same figure as Fig.~\ref{fig:NGC4147_tails} but for cluster NGC~5024. The clear overdensity in the southeastern direction is NGC~5053}
    \label{fig:NGC5024_tails}
\end{figure*}

\begin{figure*}
    \centering
    \includegraphics[width=0.9\linewidth]{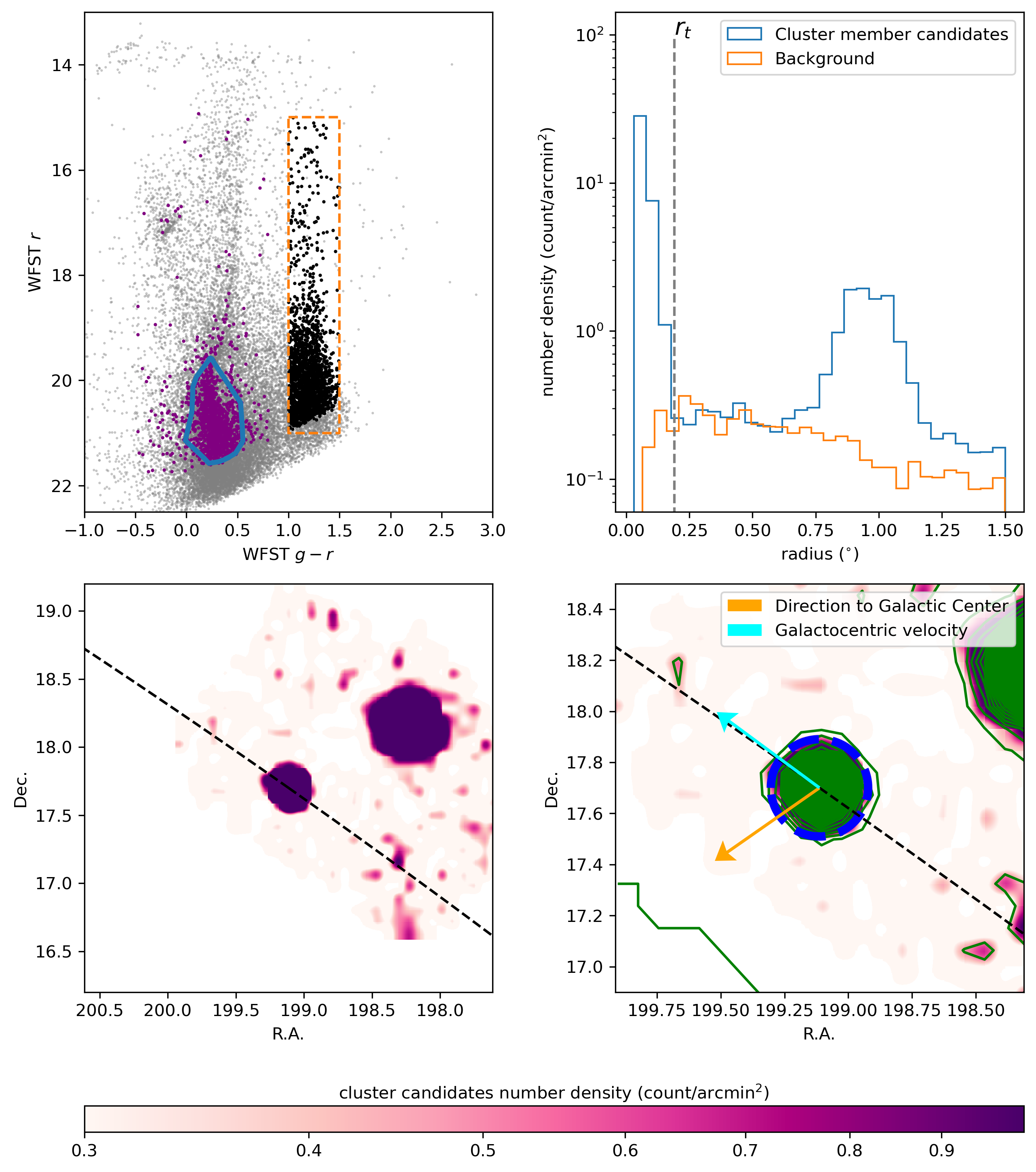}
    \caption{The same figure as Fig.~\ref{fig:NGC4147_tails} but for cluster NGC~5053. The overdensity in the northwestern is NGC~5024}
    \label{fig:NGC5053_tails}
\end{figure*}

\begin{figure*}
    \centering
    \includegraphics[width=0.9\linewidth]{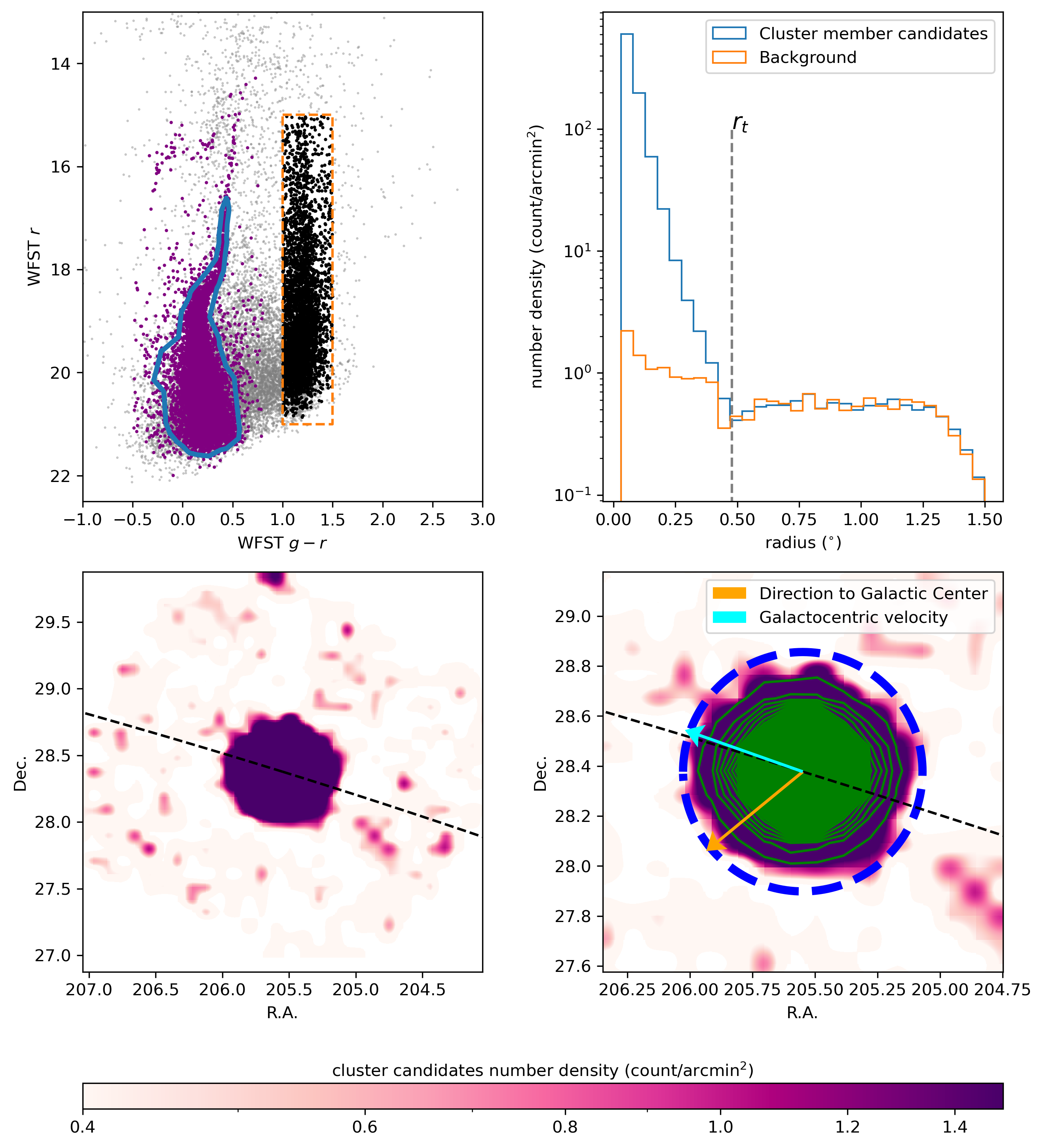}
    \caption{The same figure as Fig.~\ref{fig:NGC4147_tails} but for cluster NGC~5272}
    \label{fig:NGC5272_tails}
\end{figure*}

\begin{figure*}
    \centering
    \includegraphics[width=0.9\linewidth]{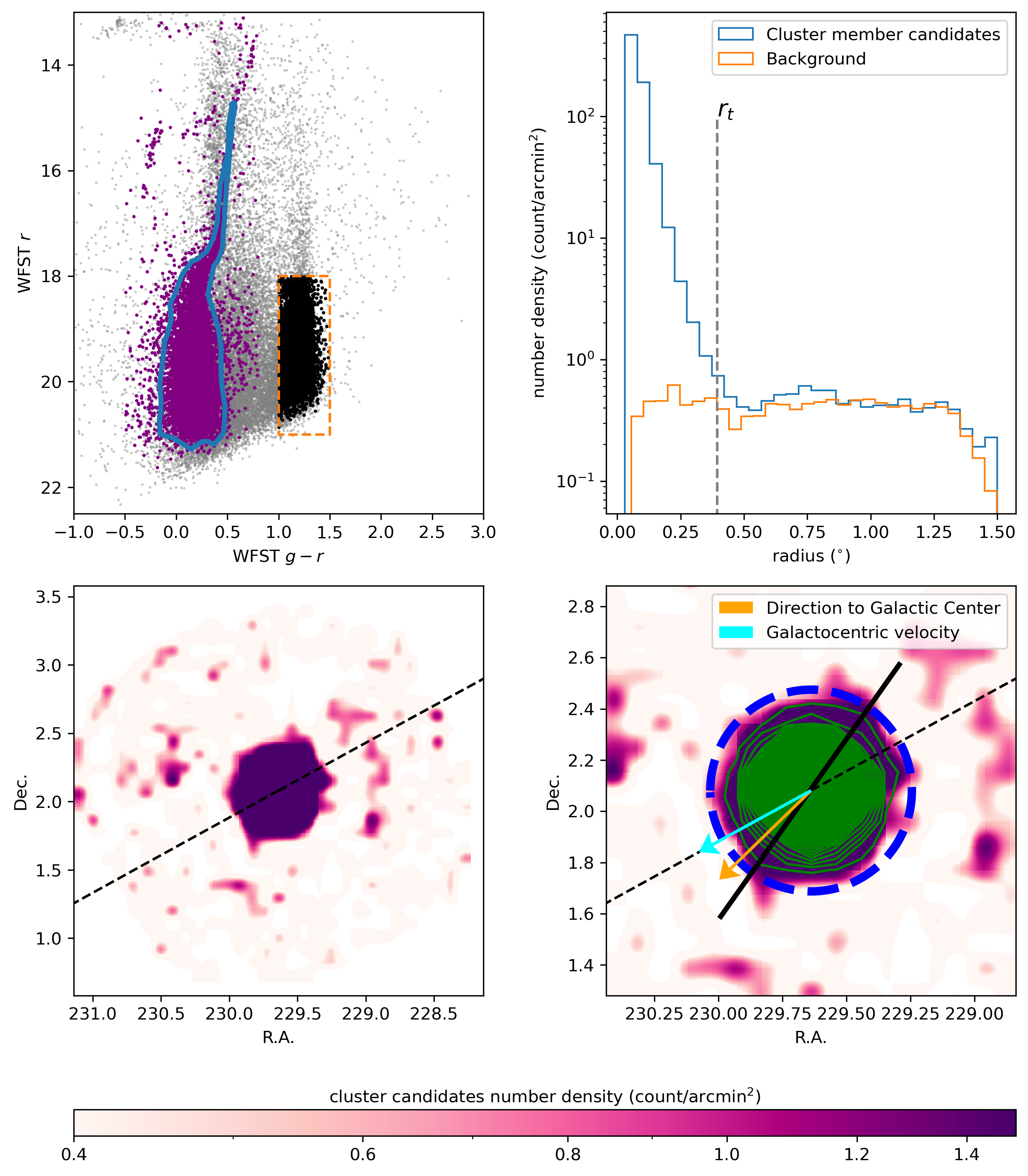}
    \caption{The same figure as Fig.~\ref{fig:NGC4147_tails} but for cluster NGC~5904. The black solid line indicate the rotation axis of this cluster measured by \citet[][]{2018ApJ...861...16L}.}
    \label{fig:NGC5904_tails}
\end{figure*}

\begin{figure*}
    \centering
    \includegraphics[width=0.9\linewidth]{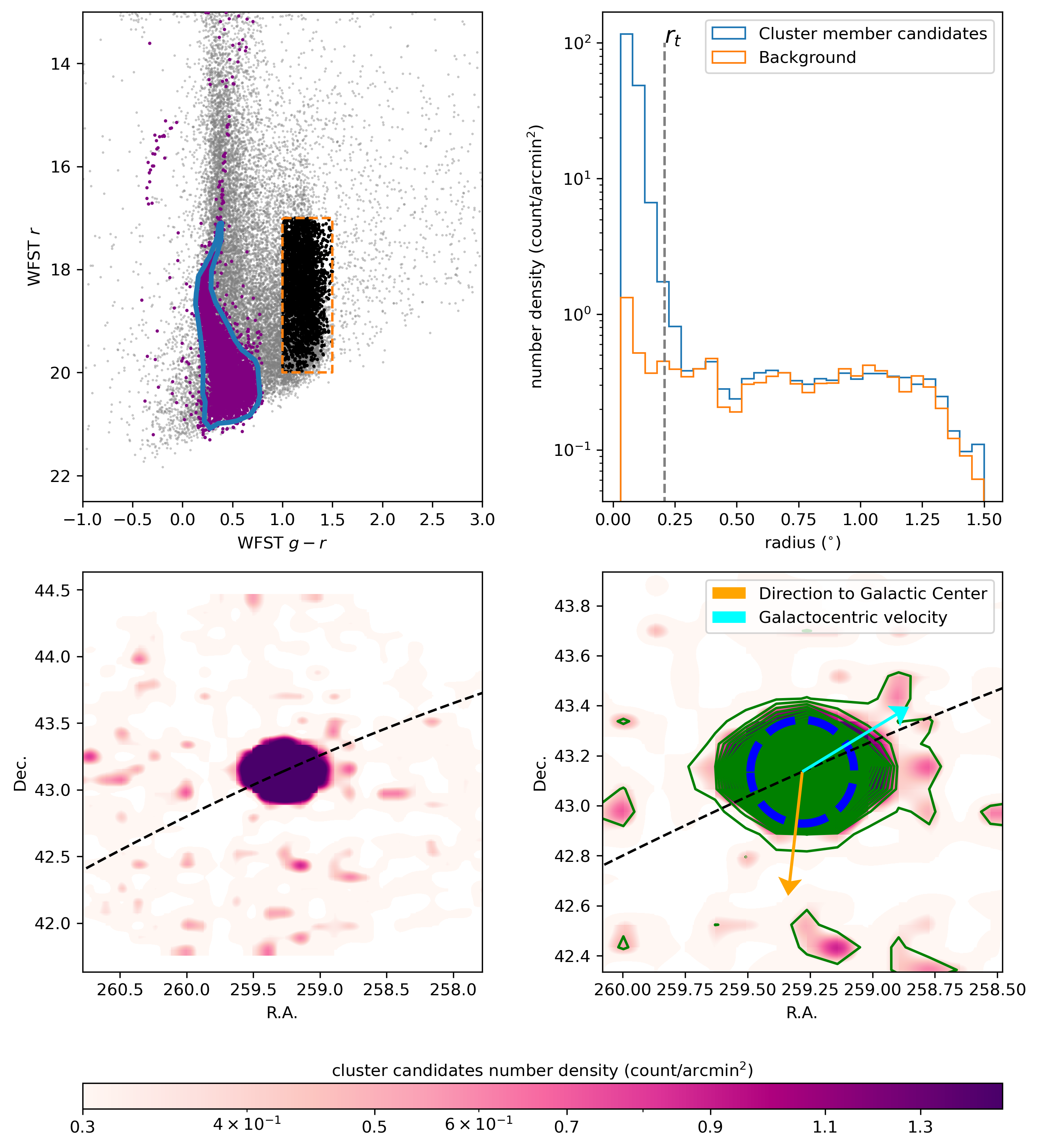}
    \caption{The same figure as Fig.~\ref{fig:NGC4147_tails} but for cluster NGC~6341}
    \label{fig:NGC6341_tails}
\end{figure*}

Using the fiducial sample as a reference, we extend the selection to identify additional member stars within a larger radius. For each star, we compute its distance in the CMD to the $i^{th}$ fiducial sample using the metric: $d = \sqrt{((g-r) - (g-r)_{i})^2 + (r - r_{i})^2}$. A star is classified as a candidate member if it has at least three closest fiducial members satisfying $d < 0.02$. For NGC~4147, this threshold is relaxed to $d < 0.08$ due to the smaller number of identified cluster members in its central region, a result of its greater distance. The blue lines in the top-right panels of Fig.~\ref{fig:NGC4147_tails}--\ref{fig:NGC6341_tails} delineate the regions from which cluster candidate stars are selected.

In addition to the member selection process, we apply the quality control Eq.~\ref{eq:quality_selection} to the selected samples. For comparison, we select a region in the CMD dominated by the Galactic thick disk, where the density of cluster members is expected to be extremely low. The top right panels of Fig.~\ref{fig:NGC4147_tails}--\ref{fig:NGC6341_tails} represent the CMD of the photometry results, where the gray points are the full photometry results. The purple points illustrate the selected samples from the central region of the corresponding clusters, displaying a clear simple stellar isochrone structure that accurately represents the location of the clusters in the CMDs. The gray points are the all targets in the field-of-view, where the black points in the orange dashed rectangle region are the selected background indicators. The top-right panels of Fig.~\ref{fig:NGC4147_tails}--\ref{fig:NGC6341_tails} display the density distribution of the selected cluster member candidates. We select stars within the orange rectangular region in the corresponding CMD as the background sample and present their density distribution in the right panels. The background number density is scaled to match the level of the cluster member density between 0.6 and 1.2 degrees from the cluster center. We further calculate the density of member stars across the field of view. Rather than presenting the raw number density of the selected member candidates, we subtract the scaled background level to estimate the neat candidates number density map. The bottom two panels illustrate the number density of the selected cluster member candidates with the background number density subtracted.

We note here that several factors can influence the member selection results. First, the results suffer from the uncertainties comes from the photometry process. The high-order CCD effects, like the responding unevenness, high-order cross-talk, and the PSF model---FWHM and the ellipticity---can vary through out the focal plane, resulting in systematic photometric uncertainties. We expect the uncertainties are included in the magnitude dispersion presented in Fig.~\ref{fig:photometric_uncertainty}, however, the detailed high precision photometric and astrometric calibration of WFST data is beyond the scope of this work, and will be published in the future WFST paper.

Another factor to consider is the differential extinction through the field of view. A matched-filter method is typically include the information from the stellar evolution isochrone. Unfortunately, at the early stage of WFST, we lack the corresponding theoretical analyses. As a compromise, we use the fiducial sample of the GC members from purely observational data with the assumption that the all the observational properties of the GC member are similar to the fiducial sample. However the differential reddening can influence the location of the location of the GC members in the CMD. Based on the measurement in previous studies \citep[e.g.,][]{bonatto_mapping_2013, 2023MNRAS.522..367L, pancino_differential_2024}, it is worth mention that the differential reddening is not significant for clusters in this study. As noted in \citet{bonatto_mapping_2013}, the differential reddening $\delta\mathrm{E(B-V)}$ of our targets are $0.019$, $0.030$, $0.029$, $0.031$, $0.033$, $0.030$, for NGC~4147, NGC~5024, NGC~5053, NGC~5272, NGC~5904 and NGC~6341 respectively, which are generally small. We note that for clusters with large differential reddening, the member selection can be significantly influenced due to the variation of the fiducial sample location and the differential reddening must be taken into consideration.

\section{Results and discussions}
\label{sec:results}

In this section, we present the observational results for each cluster. As described earlier, cluster candidates were selected from a specific region of the CMDs, defined using measurements of targets in the central region of the cluster. The background level was estimated and subtracted using stars selected from a region dominated by thick disk stars with minimal cluster members. For reference, we indicate the direction of the Galactic center based on the cluster's position and solar system position from the default "pre-v4.0" parameters from {\sc Astropy}. It includes the Galactic center of $\mathrm{(R.A., Dec.)} = (266^{\circ}.4051, -28^{\circ}.9361)$ \citep{2004ApJ...616..872R}, Galactic centre distance of $8.3$ kpc \citep{2009ApJ...692.1075G}, the Galactocentric velocity of the solar system of $(11.1, 232.24, 7.25)\ \mathrm{km/s}$ \citep{2010MNRAS.403.1829S, 2015ApJS..216...29B} and the height of the solar system to the Galactic disk of $27\ \mathrm{pc}$ \citep{2001ApJ...553..184C}. Additionally, we calculated the cluster's Galactocentric velocity using velocity information from \citet{2021MNRAS.505.5957B} . The cluster's orbit, presented in each number density figure, was derived using {\sc Galpy} \citep[][]{2015ApJS..216...29B}. Our calculated Galactocentric velocity is validated as its consistency with the tangent of the orbit.

\subsection*{NGC~4147}

This cluster has a heliocentric distance of $18.54\pm0.21\ \mathrm{kpc}$ kpc and a highly eccentric orbit ($e\sim0.87$) \citep{2021MNRAS.505.5957B}. Extended tidal arms in the outskirts of this cluster have been detected \citep{2010A&A...522A..71J}, although deeper studies have only explored one quadrant of the outskirts \citep{2012MNRAS.419...14C, 2014MNRAS.445.2971C}. During the pilot survey, we conducted 12 60-second exposures in {\it g} band and 15 60-second exposures in {\it r} band. The weather condition is not optimal during the {\it g} band exposure, and we obtained photometry of the cluster down to approximately 21.8 mag, as shown in the top-left panel of Fig.~\ref{fig:NGC4147_tails}. The top-right panel of Fig.~\ref{fig:NGC4147_tails} displays the number density distribution of cluster member candidates, along with the background number density selected from the orange rectangular region in the top-left panel. The significantly low number density at $0.5^{\circ}$ corresponds to gaps in the CCD. We mark the tidal radius from \citet[][2020 version]{1996AJ....112.1487H}. The right panel of Fig.~\ref{fig:NGC4147_tails} shows the cluster member density of NGC~4147 from our observations. The color scale indicates the number density of cluster member candidates with the background subtracted. The blue dashed circle denotes the tidal radius $r_t$ \citep[][2020 version]{1996AJ....112.1487H}, and arrows mark the direction of the Galactocentric velocity of the cluster (cyan) and the Galactic center (orange). We also denote the cluster's orbit in black dashed line.

NGC~4147 exhibits a distinct extension of cluster candidates beyond its tidal radius, reaching approximately $0.2^{\circ}$. The density map reveals a pronounced over-density in an arm-like structure extending beyond the cluster's tidal radius. This structure aligns with the findings of \citet{2010A&A...522A..71J}, who identified two tidal arms extending in the northern and southern directions. The northern arm is oriented in the direction of the cluster's Galacto-centric velocity and may extend beyond 0.5 degrees. Additionally, we find a mild signal of the multi-arm structure reported by \citet{2010A&A...522A..71J}, as evidenced by the extension of member stars in the eastern direction. Such multi-arm structures are consistent with the simulations of \citet{2007ApJ...659.1212M}, particularly for GCs at the apogalacticon of highly eccentric orbits. In addition, as the lower-right panel shows, we find a series of low-confidence overdensities of the cluster candidates along the orbit direction of the clusters, which can be a sign of the tidal tails of this cluster.

\subsection*{NGC~5024 and NGC~5053}

GC NGC~5024 has a heliocentric distance of $18.50\pm0.18\ \mathrm{kpc}$ \citep{2021MNRAS.505.5957B} and an orbital eccentricity of $\sim 0.42$. Its dynamical classification remains contentious, with proposed associations to both the Helmi Streams \citep[][]{2019A&A...630L...4M, 2020MNRAS.493..847F} and the LMS-1/Wukong accretion event \citep[][]{2022ApJ...926..107M}. While early studies reported no conclusive detection of tidal substructures \citep[e.g.,][]{2010A&A...522A..71J, 2012MNRAS.419...14C, 2014MNRAS.445.2971C}, recent claims of kinematic and spatial correlations with the Sylgr and Ravi stellar streams \citep[][]{2018ApJ...862..114S, 2019ApJ...872..152I, 2021ApJ...909L..26B} suggest the presence of large-scale tidal tails. This apparent dichotomy underscores the need for deeper photometric surveys to resolve whether the cluster exhibits weak tidal features or is embedded within a more extended stream system.

We obtained deep photometric data for this cluster with a limiting magnitude of approximately $22.3\ \mathrm{mag}$, reaching about 2 magnitudes below the main-sequence turnoff (top-left panel of Fig.~\ref{fig:NGC5024_tails}). Cluster candidates were selected using the blue-outlined CMD region, while background contamination was quantified using the orange rectangular field dominated by thick disk stars (top-right panel). The number density distribution reveals continuous stellar extensions beyond the tidal radius (gray dashed line), particularly evident in the 0.3–0.4 degree annulus.

The bottom panels display the candidate density map, showing less pronounced extra-tidal structures compared to NGC~4147's arm-like features. Nevertheless, we detect significant stellar overdensities in the southern western direction of th cluster, corresponding to the extended density enhancement around 0.6 degree. We find two high significance signal of overdensities at $\mathrm{(R.A., DEC.)} = (197.60, 18.05)\ \mathrm{and}\ (197.4, 18.00)$, located anti-aligned with the cluster's Galactocentric velocity vector. These outer overdensities exhibit spatial alignment with NGC~5024's orbital path derived from {\sc Galpy} simulations, strongly suggesting their origin as tidal debris. 

The wide-field capability of WFST proves particularly advantageous in this study, as demonstrated by the simultaneous coverage of both NGC~5024 and the southeastern overdensity structure NGC~5053 within a single exposure. With a orbit eccentricity is $\sim 0.25$ and a heliocentric distance of $17.54\pm0.23\ \mathrm{kpc}$ \citep{2021MNRAS.505.5957B}, NGC~5053 resides at a distance modulus differing by only 0.06 mag from NGC~5024, resulting in substantial photometric overlap. This photometric degeneracy introduces significant contamination from NGC~5053 members into our NGC~5024 sample, manifesting as a secondary density peak at $1^{\circ}$ offset in the top-right panel.

Contrary to the interacting cluster scenario proposed by \citet{2010AJ....139..606C}, we find no evidence of a shared tidal envelope connecting the two clusters. However, our analysis find a mild signal of NGC~5053's elongated morphology along the western axis, initially reported by \citet{2006ApJ...651L..33L} and later corroborated by \citet{2010A&A...522A..71J}. 

We performed a similar photometric and astrometric analysis on NGC~5053, with the results presented in Fig.~\ref{fig:NGC5053_tails}. A notable overdensity is detected at (R.A., Dec.) = $(197^{\circ}.75, +17^{\circ}.40)$, which aligns closely with the cluster's orbit derived from Gaia DR3 proper motions. This overdensity may represent tidal debris from NGC~5053.

\subsection*{NGC~5272}

NGC~5272 is located at a distance of $10.18 \pm 0.08$ kpc \citep{2021MNRAS.505.5957B} with an orbital eccentricity of $\sim0.50$. \citet{2021ApJ...909L..26B} proposed an association between this cluster and the Sv$\ddot{o}$l stream \citep{2019ApJ...872..152I}, suggesting the presence of extended tidal features. However, \citet{2010A&A...522A..71J} found no large-scale tidal structures, reporting only minor distortions in the outer contours. Subsequent studies by \citet{2014MNRAS.445.2971C} and \citet{2020MNRAS.495.2222S} also found no significant evidence of tidal structures in the cluster's immediate vicinity. We note that the deep photometry from \citet{2014MNRAS.445.2971C} covered only the northwestern quadrant of the cluster, potentially limiting the detection of extended features in other directions.

During the WFST pilot survey, we conducted 20 30-second {\it g}-band exposures and 25 30-second {\it r}-band exposures. Fig.~\ref{fig:NGC5272_tails} presents our deep photometric analysis of NGC~5272, reaching a limiting magnitude of approximately $22$ for main-sequence stars. The number density distribution shows no clear extension of cluster candidates beyond the tidal radius. However, we detect an intriguing overdensity feature in the southeastern direction at (R.A., Dec.) = $(204^{\circ}.7, +27^{\circ}.9)$, consisting of three aligned overdensities. While not directly connected to the cluster, the stellar density gradient suggests a possible extension toward this structure. The feature is spatially close to the cluster's orbit. In the opposite direction, we observe a slight elongation of cluster candidates, potentially indicating a symmetric tidal tail. The faintness and spatial coherence of this structure suggest it may represent a tidal feature of NGC~5272, visible only with deep photometry. If confirmed, this would provide new insights into its dynamical history.

\subsection*{NGC~5904}

NGC~5904 is located at a heliocentric distance of $7.48 \pm 0.06$ kpc and a galactocentric distance of $6.27 \pm 0.02$ \citep{2021MNRAS.505.5957B}, with an orbital eccentricity of $\sim 0.76$. As shown in the bottom-right panel of Fig.~\ref{fig:NGC5904_tails}, the cluster is currently approaching perigalacticon. Our deep photometric observations, reaching a limiting magnitude of approximately $22\ \mathrm{mag}$, cover a significant portion of the main sequence. The number density distribution in the top-right panel of Fig.~\ref{fig:NGC5904_tails} reveals clear stellar extensions beyond the tidal radius, particularly in the 0.6–0.8 degree annulus, consistent with previous findings by \citet{2014MNRAS.445.2971C} and  .

The spatial distribution (bottom panels of Fig.~\ref{fig:NGC5904_tails}) shows strong peripheral deformation, with tidal arm-like structures extending in the north and south directions. We identify several overdensities beyond the tidal radius, including two symmetric features at (R.A., Dec.) = $(230^{\circ}.6, +2^{\circ}.2)$ east and $(229^{\circ}.1, +1^{\circ}.9)$ west of the cluster. These features align with the cluster's known rotational signature \citep{2012A&A...538A..18B, 2018ApJ...861...16L, 2019MNRAS.485.1460S}. The rotational axis from \citet{2018ApJ...861...16L} (black solid line in the bottom-right panel) is nearly aligned with the north-south tidal arms and approximately perpendicular to the east-west symmetric structures, suggesting a possible connection between internal dynamics and external tidal features.

\subsection*{NGC~6341}

NGC~6341 is located at a heliocentric distance of $8.50\pm0.07$ kpc, with a galactocentric distance of $9.85 \pm 0.04$ and an orbital eccentricity of $\sim0.79$ \citep{2021MNRAS.505.5957B}. The cluster exhibits complex internal dynamics: \citet{2014ApJ...787L..26F} first reported rotational signatures, later corroborated at lower significance by \citet{2024A&A...689A.232C}, with the rotational axis aligned with the minor axis of the cluster's elliptical density distribution. Externally, large-scale tidal features have been detected, including long tidal tails reported by \citet{2020ApJ...902...89T} and clear tidal arms identified by \citet{2020MNRAS.495.2222S}.

Our observations consist of 30 exposures per filter, each with 30-second integration times, totaling 30 minutes of integration. The photometric data reach a limiting magnitude of approximately $21\ \mathrm{mag}$, enabling the detection of cluster member candidates extending beyond the tidal radius, as shown in the top-right panel of Fig.~\ref{fig:NGC6341_tails}. The spatial distribution (bottom panels) reveals a significant elliptical extension of stellar density contours beyond the tidal radius. While no high-confidence tidal structures are detected, we identify a low confidence extension of cluster candidates aligned with the cluster's Galactocentric motion vector. This feature exhibits a three-branched morphology, but its low surface brightness prevents definitive confirmation of its physical reality.

\section{Conclusions}
\label{sec:conclusion}

The 2.5-meter WFST recently completed its pilot survey, during which we conducted a series of deep imaging observations targeting five GCs in the Northern sky. This dedicated survey aims to detect and characterize tidal features in the clusters' peripheral regions by leveraging WFST's unique combination of wide field-of-view and deep photometric capabilities. Our observations reach sufficient depth to identify cluster member candidates down to the main-sequence stars, enabling the detection of low-surface-brightness features that trace the clusters' dynamical interactions with the Milky Way.

We conducted observations over five nights, targeting five distinct pointings, which included a total of six GCs. These observations covered a region with a radius of 1.5 degrees centered on each cluster. Individual exposures were stacked to achieve deep photometry, enabling precise data reduction and the derivation of CMDs for all six GCs. The limit magnitudes of the globular cluster survey exposures are about 21.5-22 magnitude, and we expect it to be 2 magnitude deeper during the formal survey. WFST shows its ability to detect faint and populated main-sequence stars, especially for NGC~4147, NGC~5024, NGC~5053 that have helio-centric distance beyond 15 kpc, which is beyond the ability of Gaia photometry of observing a significant amount of main sequence stars. 

To identify member candidates for each cluster, we selected stars located within specific regions of the CMD, defined based on the spatial distribution of targets near the cluster center. This selection criterion is functionally similar to the widely-used matched filter method. Additionally, for each cluster, we designated a control region dominated by the Galactic thick disk to estimate the background stellar density.

We compared the number density of the selected member candidates with the background density in both one-dimensional radial profiles and two-dimensional sky-projected distributions. This analysis aimed to identify potential tidal features associated with the GCs, providing insights into their dynamical evolution and interaction with the Galactic environment.

Using the number density of cluster member candidates, we investigated the spatial distribution of stars beyond the tidal radius of the cluster. For NGC~4147, the {\it g}-band observations were not conducted under optimal conditions, as the seeing exceeded 2 arcseconds, resulting in a limiting magnitude that did not reach sufficiently deep below the main-sequence turnoff. Despite these limitations, we detected prominent tidal features around the cluster.

We confirmed the presence of tidal arms in NGC~4147, which exhibit a morphology consistent with the findings of \citet{2010A&A...522A..71J}. These tidal arms extend in both the northern and southern directions, with the northern arm displaying a higher stellar density compared to the southern arm. This result aligns with the work of \citet{2022A&A...665A...8K}, who identified a similar northern overdensity using cluster candidates selected with comparable photometric depth and dynamical information. Additionally, \citet{2024AJ....168..237Z} reported a similar northern arm and an envelope-like southern overdensity using slightly deeper photometric data from the Dark Energy Survey.

We also detected a mild signal of the multi-arm structure previously observed by \citet{2010A&A...522A..71J}. Such multi-arm features have been documented in other GCs, such as NGC~288 \citep{2000A&A...359..907L} and Willman 1 \citep{2006astro.ph..3486W}. This suggests that the multi-arm structure in NGC~4147 may result from its gravitational interactions with the Milky Way. Alternatively, the structure could be linked to the host galaxy of NGC~4147. Early studies \citep[e.g.,][]{2003A&A...405..577B} proposed a spatial association between NGC~4147 and the Sagittarius (Sgr) stream, supported by radial velocity measurements and the detection of Sgr stars in the cluster's vicinity. Furthermore, main-sequence features of the Sgr stream near the cluster have been reported in several studies \citep[e.g.,][]{2004ASPC..327..255M, 2014MNRAS.445.2971C}. These findings raise the possibility that the observed extra-tidal structure may be a remnant of the Sgr dwarf galaxy population.

Along the orbital path of NGC~4147, we identified a series of low-confidence overdensities, which may represent the extension of the cluster's tidal stream. This suggests that the tidal stream aligns with the direction of the cluster's orbit. A similar alignment has been observed in Whiting 1, another GC associated with the Sagittarius (Sgr) stream \citep{2017MNRAS.467L..91C, 2022ApJ...930...23N}.

The pilot survey of NGC~4147 was conducted under non-ideal observational conditions, which limited the depth of the data to levels only marginally deeper than the main-sequence turnoff stars. Despite these limitations, the detected signals indicate that NGC~4147 likely possesses strong tidal features that remain to be fully characterized. Deeper photometric observations, such as those expected from WFST reaching beyond 24 magnitude, should be able to confirm the presence of these large-scale tidal features around the cluster.

For the clusters NGC~5024 (as well as NGC~5053) and NGC~5272, we did not detect tidal structures directly connected to the clusters. The observational depths for these clusters reached beyond 22nd magnitude, covering a significant portion of their main-sequence stars. Our results are consistent with previous studies \citep{2010A&A...522A..71J, 2014MNRAS.445.2971C, 2020MNRAS.495.2222S}. However, we identified potential overdensities around these clusters, particularly in regions close to their orbital paths. 

We note that these signals could be real or arise from fluctuations in the background contamination, which requires further confirmation from additional information. Although the photometric depths are sufficient to reliably identify a significant number of cluster candidates, additional selection criteria, such as radial velocity and proper motion measurements, would be important for confirming or refuting the existence of these structures. For instance, \citet{2023ApJ...953..130Y} employed a proper motion-weighted selection method, revealing both small-scale S-shaped tidal structures around the periphery of NGC~5272 and large-scale tidal tails extending over 50 degrees associated with the cluster. The overdensity we detected near $(\mathrm{R.A., Dec.}) = (205^{\circ}.0, 27^{\circ}.7)$ (see Fig.~\ref{fig:NGC5272_tails}) likely corresponds to a similar overdensity reported in their study. However, we did not observe the major overdensity structures near $(\mathrm{R.A., Dec.}) = (206^{\circ}.0, 27^{\circ}.5)$ presented in their work. This discrepancy may stem from differences in target selection—our analysis did not fully incorporate proper motion as a selection criterion, which could enhance the significance of the detected signals for some cases.

Similar to NGC~4147, we identified a symmetric arm-like structure in NGC~5904, extending in both the northern and southern directions. Tidal features associated with this cluster, including large-scale tidal streams \citep{2019ApJ...884..174G} and localized tidal structures around the cluster \citep{2014MNRAS.445.2971C, 2020MNRAS.495.2222S}, have been previously detected, indicating that the cluster has experienced significant tidal interactions. Interestingly, this arm-like structure is closely aligned with the rotational axis of the cluster \citep[particularly][]{2018ApJ...861...16L}. Additionally, we detected two overdensities located on either side of the rotational axis. We note that the tidal features may also contribute to the rotational signal, as the velocity difference between the two sides of the cluster could arise from a velocity gradient along the tidal arms \citep[e.g., the simulations in ][]{2021MNRAS.502.4513W}. 

The large-scale tidal structures of NGC~5904 exhibit significant complexity. \citet{2019ApJ...884..174G} detected a long trailing tail associated with this cluster, while \citet{2023MNRAS.525L..72P} proposed a dual origin for NGC~5904 based on its velocity distribution and chemical composition. They suggested that the cluster may have undergone a collision with a metal-poor GC during its evolutionary history. However, \citet{2024MNRAS.527L..32B} argued that the observed double population in metallicity could be an artifact arising from systematic errors in the data. Our findings indicate that the immediate periphery of NGC~5904 is also rich in structural features, further highlighting the dynamical complexity of this cluster. These results underscore the need for continued investigation into the tidal interactions and evolutionary history of NGC~5904.

NGC~6341 stands out as a unique GC among our targets. The number density distribution of its member candidates exhibits a clear deviation from a circular shape, displaying a symmetric elliptical contour beyond the tidal radius. This elliptical morphology may result from internal dynamical processes. For instance, the minor axis of the ellipse aligns with the rotational signal of the cluster \citep{2024A&A...689A.232C}. If the shape is indeed linked to the cluster's rotation, stars in the corresponding region should remain bound to the cluster and exhibit systemic rotation.

Alternatively, the elliptical shape could represent an extension of the cluster's tidal arms. NGC~6341 has been shown to possess strong tidal features, as recently documented by \citet{2020MNRAS.495.2222S, 2021ApJ...914..123I}, and a large-scale tidal stream identified by \citet{2020ApJ...902...89T}. In the leading direction of the cluster's motion, we detected low-confidence overdensities extending from the cluster, suggesting that stars at the periphery may no longer be gravitationally bound. Additionally, the observed rotational signal could arise from a velocity gradient along the tidal arms of the cluster. To resolve these possibilities, future studies should focus on dynamical analyses of stars in this region and the development of detailed N-body models to better understand the cluster's evolution and tidal interactions.

The wide field survey on GCs is essential for the precise GC population estimation \citep[e.g.,][]{Sarajedini_2007}, multi-population studies \citep[e.g.,][]{2013MNRAS.431.2126M, 2015AJ....149...91P, 2019MNRAS.485.3042S,2023MNRAS.522.2429M}, and the stellar distribution \citep[e.g.,][]{2019MNRAS.485.3042S}. This study demonstrates the capability of WFST in achieving deep photometric observations. We expect WFST can provide another important contribution with its $u$-band photometry in its future survey for studies on the multiple population studies. More importantly, WFST is promising for the interdisciplinary studies combining the variables stars and the GC properties with the time-series observations, focusing on the targets including the blue straggler stars and the binary populations. Hence, the future survey campaigns and subsequent data releases hold significant promise for advancing research in near-field cosmology. 

\section*{Acknowledgements}
The Wide Field Survey Telescope (WFST) is a joint facility of the University of Science and Technology of China, Purple Mountain Observatory. This work is supported by National Key Research and Development Program of China (2023YFA1608100) , the National Natural Science Foundation of China (NSFC, grant No. 12173037, 12233008), the CAS Project for Young Scientists in Basic Research (No. YSBR-092), the Fundamental Research Funds for the Central Universities (WK3440000006) and Cyrus Chun Ying Tang Foundations.
ZW acknowledge the support of the Youth Innovation Fund (WK2030000080); China Postdoctoral Science Foundation (2022M723059); China Postdoctoral Science Foundation (2023T160615). XZL acknowledge the support from Anhui Provincial Natural Science Foundation (2308085QA35).
\section*{Data Availability}
The data collected by WFST will be made publicly available in future data releases. At present, the ongoing survey data are restricted to access within the collaboration. For detailed information and access to these data, please reach out to the collaboration (https://wfst.ustc.edu.cn).
In addition, the analysis results, including the density map, are available upon query.



\bibliographystyle{mnras}
\bibliography{ref} 







\bsp	
\label{lastpage}
\end{document}